**Prospects for non-linear memristors as so-far missing core hardware element for transfer-less data computing and storage**

Heidemarie Schmidt[1,2;3*]

[1]*Department Quantum Detection, Leibniz Institute of Photonic Technology, Albert-Einstein-Str. 9, 07745 Jena, Germany*

[2]*Institute for Solid State Physics, Friedrich Schiller University Jena, Helmholtzweg 3, 07743 Jena, Germany*

[3]*TECHiFAB GmbH, Bautzner Landstraße 45 01454 Radeberg, Germany*

*\*Corresponding author: Heidemarie Schmidt (Heidemarie.Schmidt@uni-jena.de)*

**Abstract**

We like and need Information and Communications Technologies (ICT) for data processing. This is measureable in the exponential growth of data processed by ICT, e.g. ICT for cryptocurrency mining and search engines. So far, the energy demand for computing technology has increased by a factor of 1.38 every ten years due to the exponentially increasing use of ICT systems as computing devices. The energy consumption of ICT systems is expected to rise from 1500 TWh (8% of global electricity consumption) in 2010 to 5700 TWh (14% of global electricity consumption) in 2030 [1]. A large part of this energy is required for the continuous data transfer between the separated memory and processor units which constitute the main components of ICT computing devices in von-Neumann architecture. This at the same time massively slows down the computing power of ICT systems in the von-Neumann architecture. In addition, due to the increasing complexity of AI compute algorithms, since 2010 the AI training compute time demand for computing technology increases tenfold every year, for example in the period from 2010 to 2020 from $1*10^{-6}$ to $1*10^{+4}$ Petaflops/Day [2]. It has been theoretically predicted that ICT systems in the neuromorphic computer architecture will circumvent all of this through the use of merged memory and processor units. However, the core hardware element for this has not yet been realized so far. In this work we discuss the prespectives for non-linear resistive switches as the core hardware element for merged memory and processor units in neuromorphic computers.

**Overview of the topic**

Largest part of the energy consumed by ICT systems in von-Neumann architecture is required for the continuous data transfer between the separated memory and processor units [1,2]. First ICT systems in von-Neumann architecture consisted of vacuum tubes as processor unit and magnetic core memory as memory unit (Fig. 1A). Our current ICT systems in von-Neumann architecture consist of transistors as processor unit and random access memories (RAM) as memory unit (Fig. 1B). The theoretically predicted ICT system in neuromorphic computer architecture merges the processor unit and the memory unit (Fig. 1C) and is expected to process and store data transferless. In this work we discuss the perspectives of non-linear memristors as the core hardware element for neuromorphic computer architecture.

First we will have a closer look at the transport characteristics, e.g. current-voltage curves, of processor units in von-Neumann computers, namely of vacuum tubes (Fig. 1A) and transistors (Fig. 1B). Because for a linear signal input the signal output of such processor units is strongly distinguishable, they are used as basic building blocks in the arithmetic logic unit of von-Neumann computers. The I-V curves of such processor units are non-linear and pass through the origin, implying that they do not primarily store energy. However, the signal output is lost when the operation power of such processor units is switched off before storing the signal output in the memory unit. Therefore, the requirements for the core hardware element for merged memory and processor units in neuromorphic computers are non-linear current-voltage curves and the storage of the signal output as internal state parameter in this core hardware element (Fig. 2C). The material, $BiFeO_3$ (BFO), for such a core hardware element, as been found in 2011 [3] and storage of the signal output as internal state parameter in a BFO-based core hardware element has been reported in 2014 [4].

The BFO-based core hardware element belongs to the memristor class of passive hardware elements. A memristor is a passive electronic component that has the ability to change its resistance in response to an applied voltage (voltage-driven) or in response to the current that flowed through the component (current-driven) [5]. In this work we focus on memristors which are commonly referred to as resistive random access memory (ReRAM) devices.

The purpose of this article is to discuss the transport properties of different memristors to better understand why memristors with linear I-V characteristic curves can only be used as memory units and why memristors with non-linear I-V characteristic curves can be used as both, as procssor and as memory unit without data transfer during computation. Typically, memristors are reconfigurable resistive switches are composed of a poorly conducting dielectric thin film with a conducting metallic top electrode and a bottom electrode attached. So far available reconfigurable resistive switches only enable data storage, but no data processing in the same device. Only memristors with non-linear I-V curves [3] have been shown to be able to process and store data in the same device [4], thus fulfilling the requirements of the so far missing core hardware for resource-saving neuromorphic computers.

From now on we differentiate between linear memristors (Fig. 2 A-C) and non-linear memristors (Fig. 2D). Strukov et al. discovered the $TiO_2$-based linear memristor [6]. Since then many other groups reported on different types of linear memristors [7-19]. Up to our knowledge there was the first report on a non-linear memristor, on the BFO-based memristor, in 2011 [3]. Only recently, another type of non-linear memristor, the two-dimensional $MoS_2$-based memristor, has been presented [20]. However, reported infinitely large initialization time of the internal state parameter [20] will limit the use of such non-linear, two-dimensional memristors as a core hardware element in merged memory and processor units in neuromorphic computers. The initialization of the non-linear, BFO-based memristor [3] can be ultrashort [21] and has been recently described by a physical model [22]. We discuss the effect of initialization on internal state parameters later (Fg. 3D).

The central difference between linear and non-linear memristors can be understood from the four criteria derived by Prof. Leon O. Chua [5] to describe the transport properties of memristors:

1. The electrical transport properties of the memristor are characterized using current-voltage measurements.

2. The current-voltage measurements show current-voltage curves like those of a diode characteristic, but with a hysteresis.

3. The memory of a memristor is reflected in a hysteresis of the current-voltage curves. The greater the hysteresis of the current-voltage curve, the greater the memory of a memristor.

4. The state of a memristor (memristivity or memristance M) clearly depends on the change in charge q when sweeping the current or on the change in flux $\phi$ when sweeping the voltage.

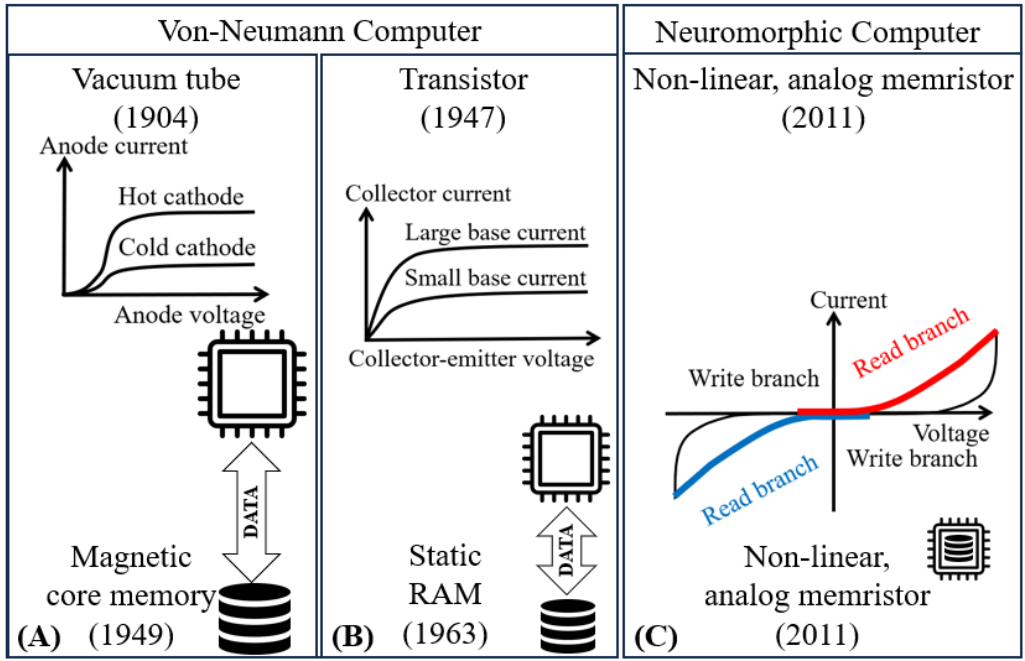

Fig. 1. Processor and memory unit in (A,B) von-Neumann computers and in Neuromorphic computers with (A,B) separated processor and memory component in von-Neumann architecture and with (C) merged processor and memory unit in Neuromorphic architecture. The data transfer between the processor and memory in von-Neumann computers is resource-consuming.

The central difference between linear and non-linear memristors prelies in the one-to-one dependency of the memristance (charge or flow) on the previous change in current or voltage. Formally, this corresponds to the 4$^{th}$ criterion in Prof. Chua's definition of a memristor. In the following memristors that only meet the first three criteria (1-3) are referred to as linear memristors and memristors that meet all four criteria (1-4) are referred to as non-linear memristors. Two behavioral characteristics of non-linear memristors can clearly distinguish them from linear memristors:

1$^{st}$ property: The resistance of a non-linear memristor is characterized by successive, analog changes and not by abrupt, digital changes

A non-linear memristor can assume all resistance values between the high resistance (HRS) and the low resistance (LRS) in a non-volatile manner, regardless of whether the resistance value of the memristor is changed from the HRS to the LRS or from the LRS to the HRS. First linear memristors were linear filamentary memristors which assumed a maximum HRS value and a minimum LRS value (Fig. 2B). Nowadays new types of linear memristors with many additional resistance values between maximum HRS value and a minimum LRS value have beend developed [18,19]. Please note that recently also linear memristors with many resistance values between HRS and

If linear memristors are to be used as artificial synapses in neuromorphic chips, this non-ideality must be compensated as best as possible by additional software and even hardware. However, this is only partially successful. Non-linear memristor-based artificial synapses are ideal and do not require additional software and hardware to compensate non-idealities.

$2^{nd}$ property: The storage and processing of data takes place in one and the same non-linear memristor device cell

Typically, memristors are reconfigured above a current or voltage threshold value and the state of the resistive switch is read out far below a current or voltage threshold value, e.g. at the red and blue semicircles shown in Fig. 2. Compared to linear memristors with (B) filamentary and (C) structural switching the non-linear memristors do not have such a threshold, i.e. the state of the non-linear memristor can be read out at any current or voltage value on one of the two read branches and in the small current or bias range of the two write branches of the hysteretic current-voltage (I-V) curves [22].

In the following we analyze the I-V characteristics of different memristors also with regard to their ability to store and process data in the same memristor cell using cycled hysteretic current-voltage curves where the voltage is linearly ramped. This includes different types of linear memristors such as ferroelectric (Fig. 2A), filamentary (Fig. 2B), and phase-change (Fig. 2C) memristors. Typically, linear memristors are reconfigured above a current or voltage threshold value and the state of the linear memristor is read out below a current or voltage threshold value (Fig. 2 A-C). Linear memristors are not suitable as core hardware components for the merged memory and processor unit in neuromorphic computers, since they can only store but not process data. Compared to linear memristors non-linear memristors do not have such a reconfiguration threshold. As will be explained in the following non-linear memristors have dynamic and static internal state variables in the two full write branches and in the two full read branches, respectively. This uniquely allows the readout of the resistance state of the non-linear memristor in the two complete read branches of the current-voltage characteristics (Branch 2 and Branch 4 in Fig. 2D). Understanding this allows to operate non-linear memristors as memory and processor in the same cell without data transfer and – as a long-term goal – to implement such non-linear memristors as core hardware element in resource-efficient neuromorphic computers. This might also open – as a short-term goal – new avenues for performance improvement of devices in edge computing, edge sensorics, and secure electronics. The switching behaviour of reconfigurable resistive switches with a conducting metallic top electrode and a bottom electrode is typically analyzed by current-voltage measurements.

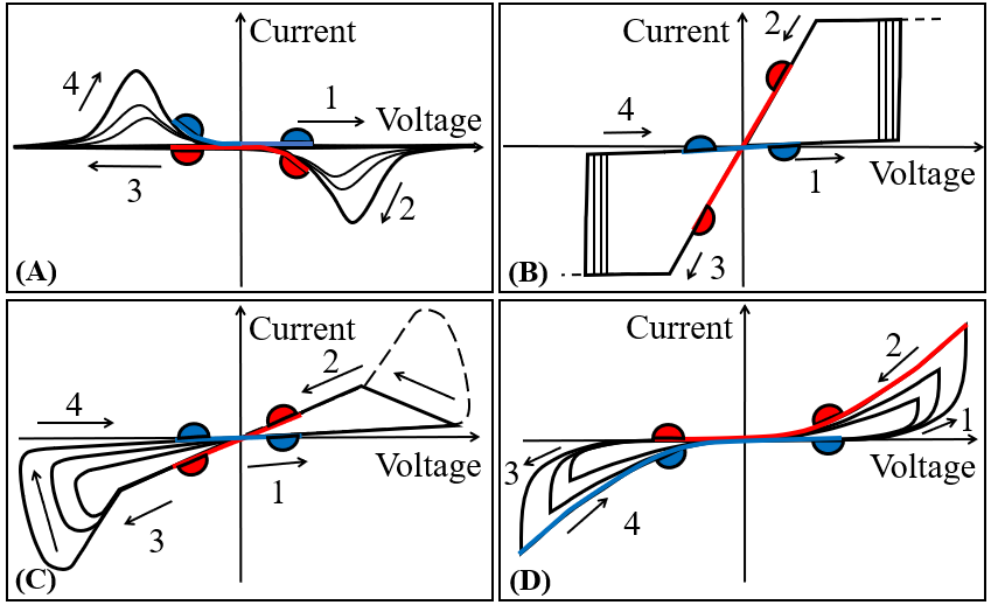

Fig. 2. Voltage sourced hysteretic current-voltage (I-V) characteristic curves of memristors with (A) ferroelectric, (B) filamentary, (C) structural, and (D) barrier switching. (D) Only I-V curves of memristor with barrier switching touch at the origin of the I-V coordinate system, i.e. I=I(0V)=0A. The I-V curves of memristors with (B) filamentary switching and with (C) structural switching are crossed and the I-V curves of memristors with (A) ferroelectric switching overlap in a broader range around the origin of the coordinate origin. Branches of the current-voltage curves are numbered 1 through 4. Arrows indicate the sequence of voltages sourced: Branch 1: 0→+$V_{max}$, Branch 2: +$V_{max}$→0, Branch 3: 0→-$V_{max}$, Branch 4: -$V_{max}$→0. Segments of the current-voltage curves where the internal state variables of the memristor do not change are shown in red and blue.

In the following we analyze the possible operation of linear and non-linear memristors. The current-voltage curves can be recorded in both, in current-driven and voltage-driven-mode. For example, in the voltage-driven mode the voltage can be applied to the top electrode and the bottom electrode can be grounded. Then the voltage is changed successively. For convenience, we will number the branches on the current-voltage curve of memristors (Fig. 2), e.g. branch 1 from 0 V to $+V_{max}$, branch 2 from $+V_{max}$ back to 0 V, branch 3 from 0 V to $-V_{max}$, and branch 4 from $-V_{max}$ back to 0. The current-voltage curve of a memristor exhibits an hysteresis in the positive voltage range, if the current in branch 1 and the current in branch 2 which are measured at the same positive voltage are different. And the current-voltage curve of the memristor exhibits an hysteresis in the negative voltage range, if the current in branch 3 and the current in branch 4 which are measured at the same negative voltage are different. If the current-voltage characteristic remains unchanged after repeated cycling, this is a first indication of a high endurance of the memristive switches. Reliable endurance measurement methods [23] with fast initialization of equilibrium resistance state will accelerate the integration of reconfigurable resistive switches in commercial products.

When storing data, memristors differ in their switching behavior (see Fig. 2A-C). If an appropriate write voltage is applied, then the internal state variables of ferroelectric (Fig. 2A), filamentary (Fig. 2B), and structural (Fig. 2C) linear memristors are constant and current-voltage characteristics can be described either using branches 4 and 1 or branches 2 and 3. The resistance state of the linear memristors is read out by applying a read voltage. The read voltage and the corresponding read current in branches 4 and 1 and in branches 2 and 3 is represented exemplarily by blue and red semicircles, respectively (Fig. 2). The read voltage must be selected so small that the internal state parameter is not changed when reading. The range of possible read voltages and corresponding read currents in branch 4 and 1 and in branch 2 and 3 is indicated by blue and red lines, respectively. Interestingly, for a ferroelectric linear memristor (Fig. 2A) and for a barrier non-linear memristor (Fig. 2D) the read current depends on the polarity of write and read current. For the ferroelectric linear memristor (Fig. 2A), the read voltage must be chosen to be quite large, since no current flows with a small read voltage around 0 V. Therefore, every readout results in a small change of internal state variables of the ferroelectric linear memristor (so-called read-disturbs) and a read disturb-aware write operation has to be applied. For the barrier non-linear memristor (Fig 2A) the read voltage in branch 1 and 3 is chosen typically small, however the read voltage in branch 2 and 4 can be chosen over the full voltage range without changing internal state variables.

In a digital memory it is sufficient to store two states, high resistance state (HRS) and low resistance state (LRS). The states of such digital memories are retained until a write voltage is applied again, thus changing the state of the memory. For example, if a positive and a negative read voltage is selected and if the linear memristor is in HRS, then the read current is small (blue semicircle on branch 1 and blue semicircle on branch 4 in Fig. 2 B-C). If a positive and a negative read voltage is selected and if the linear memristor is in LRS, then the read current is large (red semicircle on branch 2 and red semicircle on branch 3 in Fig. 2 B-C). On the other hand, if a positive and a negative read voltage is selected and if the non-linear memristor is in branch 4 or branch 1, then the read current is large (blue semicircle on branch 4 in Fig. 2 D) or the read current is small (black semicircle in branch 1 in Fig. 2D). If a positive and a negative read voltage is selected and if the non-linear memristor is in branch 2 or branch 3, then the read current is large (red semicircle on branch 2 in Fig. 2 D) or the read current is small (black semicircle in branch 3 in Fig. 2D).

**Own contribution to the field**

In the evaluation as a resistive random access memory (ReRAM) device, attention is first directed to how large the difference between the read current and thus the resistance ratio between the resistance in the HRS and the resistance in the LRS is. Programming cycles $\geq 0.5 \cdot 10^6$ can be used industrially at least in the temperature range between 0 – 80°C. The retention describes how long the HRS state and LRS state can be kept by the memristive switch. Values $\geq 10$ years at least in the temperature range between 0 – 80°C can be used industrially. Established retention and endurance measurements and a first analysis of the current-voltage curve of a resistive switch yields the benchmark parameters for a comparison with commercially available resistive switches, e.g. of Resistive Random Access Memory (ReRAM) devices which store information by changing the electrical resistance of a poorly conductive dielectric between top electrode and bottom electrode. If the ReRAM has its maximal resistance $R_{OFF}$, it is in high resistance state (HRS) and if the ReRAM has its minimal resistance $R_{ON}$, it is in low resistance state (LRS). For classic ReRAM components, $R_{OFF}/R_{ON}$ should be greater than 10, retention should be larger than 10 years and endurance larger than $10^6$. The $R_{OFF}/R_{ON}$ ratio of ferroelectric linear memristors

($R_{OFF}/R_{ON}=10^0$ .. $10^1$) [8, 16-18], is smaller then the $R_{OFF}/R_{ON}$ ratio of structural linear memristors ($R_{OFF}/R_{ON}=10^2$..$10^3$) [11-14] and of filamentary linear memristors ($R_{OFF}/R_{ON}=10^2$.. $10^5$) [7,9,10,19]. The $R_{OFF}/R_{ON}$ ratio of non-linear memristors ($R_{OFF}/R_{ON}=10^1$.. $10^3$) [20, 24-32] is comparable to the $R_{OFF}/R_{ON}$ ratio of filamentary and structural linear memristors. Only for the non-linear memristor does the read current depend on the polarity of the write voltage and on the polarity of the read voltage. On the other hand, the read current of a linear memristor (except ferroelectric linear memristor with read disturbs) depends only on the polarity of the write voltage and not on the polarity of the read voltage.

(1) A first important feature of a non-linear memristor is certainly the programming with positive and negative write voltage, but in particular different read currents with positive and negative read voltage: This is also called barrier switching which enables the operation as one of the 16 Boolean logic gates, since two variables are available for implementation with positive and negative read voltage [4].

(2) Another important feature is the write voltage range in which the internal state variables of a memristor can be programmed to the maximum and minimum read current. The larger the corresponding read current range, the more internal state variables can be stored.

(3) Filamentary or structural linear memristors do not fulfill feature (1), which can be seen from the crossing hysteretic I-V curves (Fig. 2 B-C) in comparison to tangential hysteretic I-V curves (Fig. 2 D).

Finally, we take Chua's work literally [5] and analyze the memristance of linear memristors [6] and of non-linear memristors [33]. As an example for linear memristors we take the $TiO_2$-based, linear memristor [6] and as an example for non-linear memristor we take the BFO-based memristor [3]. To keep the discussion of linear and non-linear memristor comparable, we ramp the voltage and measure the current. Furthermore, we label the branches of corresponding current-voltage characteristics as follows: for the voltage ramped from 0 V to +$V_{max}$ "Branch 1", ramped back from +$V_{max}$ to 0 V "Branch 2", ramped from 0 V to -$V_{max}$ "Branch 3", and ramped from -$V_{max}$ to 0 V "Branch 4". The current-voltage characteristics and labeled branches of linear and non-linear memristor are sown in Fig. 3A and in Fig. 3B, respectively. In the small bias range the resistance of the linear memristor does not change and can be used to distinguish the linear memristor being in HRS (branch 4 and branch 1 in Fig. 3A) or being in LRS (branch 2 and branch 3 in Fig. 3A). In the large bias range the resistance of the linear memristor continuously changes (branches 1-4 in Fig. 3A). On the other hand the resistance of the non-linear memristor continuously changes in the small and in the large bias range (branches 1-4 in Fig. 3B). From that one can conclude that a change of resistance R is not directly related with a change of the internal state variables of a memristor. Only a change of memristance M [33] which is the derivative of flux $\phi$ with respect to charge q [5], is directly related with a change of the internal state variables. To determine the memristance M from flux-charge characteristic curves, we integrate the sourced voltage V=V(t) and the measured current I=I(t) of the hysteretic current-voltage curve of a linear (Fig. 3A) and of a non-linear (Fig. 3B) memristor as follows:

$$\varphi = \int_{-\infty}^{t} V(t)dt = \int_{-\infty}^{t_0} V(t)dt + \int_{t_0}^{t} V(t)dt = \varphi(t_0) + \int_{t_0}^{t} V(t)dt, \tag{1}$$

$$q = \int_{-\infty}^{t} I(t)dt = \int_{-\infty}^{t_0} I(t)dt + \int_{t_0}^{t} I(t)dt = q(t_0) + \int_{t_0}^{t} I(t)dt, \tag{2}$$

where flux ϕ results from the integration of time dependent sourced voltage V=V(t) and where charge q results from integration of time dependent measured current I=I(t). Here we introduce an offset $\varphi(t_0)$ and $q(t_0)$ which can be defined during the initialization step. After initialization the non-linear memristor will reveal well-defined internal state variables. We plotted the flux ϕ vs. charge q of the linear, TiO$_2$-based memristor (Fig. 3C) and of the non-linear, BFO-based memristor (Fig. 3D) and mark the offset $\varphi(t_0)$ and $q(t_0)$ after the initialization step with an open circle (Figs. 3C and 3D). The flux controlled (bias sourced) memristance is given as follows:

$$M(\varphi(t)) = \frac{d\varphi(t)}{dq(t)}. \tag{3}$$

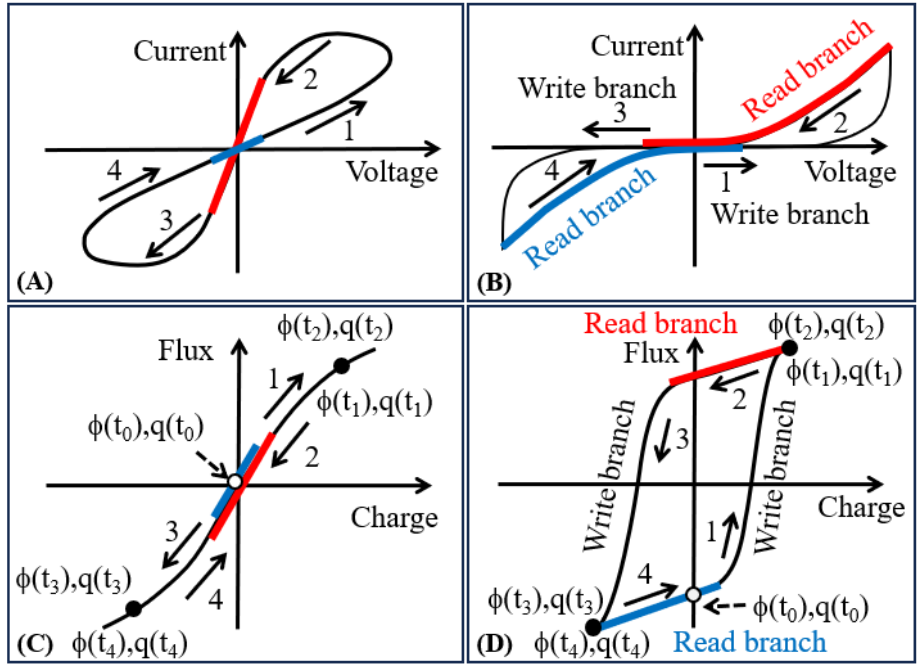

Fig. 3. Schematic representation of (A,B) current-voltage (I-V) characteristic curves and of (C,D) flux-charge ($\Phi$-Q) characteristic curves of a (A,C) linear memristor and of a (B,D) non-linear memristor. (A,B) The current-voltage curves of both memristors are hysterestic. The flux-charge curve of the (C) linear memristor is non-hysteretic and of the (D) non-linear memristor is hysteretic. (B,D) Segments of the current-voltage curves where the internal state variables of the memristor do not change are shown in red and blue. Arrows indicate the sequence of voltages sourced: Branch 1: $0 \rightarrow +V_{max}$, Branch 2: $+V_{max} \rightarrow 0$, Branch 3: $0 \rightarrow -V_{max}$, Branch 4: $-V_{max} \rightarrow 0$. (B,D) Segments of the current-voltage curves where the internal state variables of the memristor do not change are shown in red and blue. (D) We suggest to use initialization until time point $t=t_0$ to define the internal state variables before sourcing the voltage $0 \rightarrow +V_{max} \rightarrow 0 \rightarrow -V_{max} \rightarrow 0$ (branches 1-4).

For completeness we also indicate the charge controlled memristance which is given as follows:

$$M(q(t)) = \frac{d\varphi(t)}{dq(t)}. \tag{4}$$

The flux-charge curve of the linear memristor is non-hysteretic (Fig. 3C). Its slope is constant in the small bias range and its slope changes continuously in the large bias range. That means that the internal state variables of a linear memristor continuously change in the large bias range. For every time point on the time scale t=t$_1$ on branch 1 and memristance M(ϕ(t$_1$)), another time point t=t$_2$ on branch 2 and memristance M(ϕ(t$_2$)) can be found for which applies M(ϕ(t$_1$))=M(ϕ(t$_2$)). And for every time point on the time scale t=t$_3$ on branch 3 and memristance M(ϕ(t$_3$)), another time point t=t$_4$ on branch 4 and memristance M(ϕ(t$_4$)) can be found for which applies M(ϕ(t$_3$))=M(ϕ(t$_4$)). From that follows that for no time point it is possible to distinguish between the internal state variables of the linear memristor with memristance from branch 1 and with memristance from branch 2 or between the internal state variables of the linear memristor with memristance from branch 3 and with memristance from branch 4 (Fig. 3 C).

On the other hand, the flux-charge curve of the non-linear memristor is hysteretic (Fig. 3D) and changes its slope continuously in the two write branches, namely in branch 1 and in branch 3, and has a constant slope in the two read branches in the small and in the large bias range.

Only for a single time point on the time scale $t=t_1$ on branch 1 and memristance $M(q(t_1), \phi(t_1))$, another time point $t=t_2$ on branch 2 and memristance $M(\phi(t_2))$ can be found for which applies $M(q(t_1), \phi(t_1))=M(q(t_2), \phi(t_2))$. This is at $V(t=t_1=t_2)=+V_{max}$ (Fig. 3D). And also only for a single time point on the time scale $t=t_3$ on branch 3 and memristance $M(q(t_3), \phi(t_3))$ another time point $t=t_4$ on branch 4 and memristance $M(q(t_4), \phi(t_4))$ can be found for which applies $M(q(t_3), \phi(t_3))=M(q(t_4), \phi(t_4))$. This is at $V(t=t_3=t_4)=-V_{max}$ (Fig. 3D). From that follows that for every time point, except at the time point $t=t_1=t_2$ and at the time point $t=t_3=t_4$, it is possible to distinguish between the internal state variables of the non-linear memristor (Fig. 3 D). Furthermore, because the memristance is constant in branch 2 and in branch 4, we call branch 2 and branch 4 "read branches". And because the memristance is continuously changing in branch 1 and in branch 3, we call branch 1 and branch 3 "write branches". We used this concept to extract internal state variables from current-voltage curves of a non-linear memristor [22]. We assumed that the internal state variables are static and dynamic in the two full read branches and in the two full write branches, respectively. Furthermore, extracted internal state variables are the same at the time point $t=t_1$ on the write branch 1 and at the time point $t=t_2$ on the read branch 2 with $V(t=t_1=t_2)=+V_{max}$ where memristance is the same, i.e. $M(q(t_1), \phi(t_1))=M(q(t_2), \phi(t_2))$, and extracted internal state variables are the same at the time point $t=t_3$ on the write branch 3 and at the time point $t=t_4$ on the read branch 4 with $V(t=t_3=t_4)=-V_{max}$ where memristance is the same, i.e. $M(q(t_3), \phi(t_3))=M(q(t_4), \phi(t_4))$. The static and dynamic internal state variables of a non-linear memristors can be used to implement operations on non-linear memristors representing linear, non-linear, and even transcendental, e.g. exponential or logarithmic, input-output functions.

The future implication of presented findings is closely linked to the computability of a problem, i.e. to the existence of an algorithm to compute input-output maps of the computable functions in finite time. According to Boche, Fono, and Kutyniok [34,35] computing and processing paradigm needs to change so that more efficient but equally powerful computations can be carried out. In computational theory, the Turing machine is a computational model to describe calculations over a (finite) rational number of states with unlimited memory and time at its disposal and the Blum-Shub-Smale (BSS) machine is a computational model to describe calculations over an (uncountable) real number of states with registers that can store arbitrary real numbers and that can compute rational functions over reals in a single time step. So far, there exists no computational models to describe calculations over (uncountable) real numbers of states with registers that can store arbitrary real numbers and that can compute transcendental, e.g. exponential or logarithmic, functions over reals in a single time step. Because non-linear memristor devices are expected to perform computing of input-output maps of transcendental functions in a single time step without intermediate data storage, it is proposed to incorporate non-linear memristor devices into the computing pipeline as registers. The memristance of the non-linear memristor-based registers can be written in the write branches 1 and 3 and read in the read branches 2 and 4.

To summarize, non-linear memristors are promising as the so-far missing core hardware for ressource-saving neuromorphic computers. Their behavior is similar to that of a synapse in the brain, a component with memory value. Novel microelectronic circuits with non-linear memristors will allow for digital and analog data processing in real-time and in a certifiable manner in the same device without data transfer, whereby transparent, repeatable, and trustworthy data processing algorithms are performed in hardware. As an exemplarily embodiment for non-linear memristors we analyzed the hysteretic current-voltage (I-V) and hysteretic flux-charge ($\phi$-q) of the $BiFeO_3$ (BFO)-based, non-linear memristor [3] and compared it with the hysteretic current-voltage (I-V) and non-hysteretic flux-charge ($\phi$-q) of the $TiO_2$-based, linear memristors [6]. The BFO-based, non-linear memristor has two write branches and two read branches with dynamic and static internal state variables, respectively. Only the non-linear memristor fulfills the requirements of the core hardware for resource-saving neuromorphic computers. Furthermore, because non-linear memristor devices are expected to perform computing of input-output maps of transcendental functions in a single time step, it is proposed to incorporate non-linear memristor devices as registers in neuromorphic computers that can compute transcendental, e.g. exponential or logarithmic, functions over reals in a single time step.

## ACKNOWLEDGMENTS

I wish to thank my former and actual team members who continuously developed the BFO-based, non-linear memristor for edge computing, edge sensorics, and secure electronics. For this work I mainly wish to thank Prof. Bernd Ulmann from FOM Hochschule für Oekonomie & Management for motivating the development of nonlinear analog hardware for analog computers, Dr. Thomas Wille and Dr. Ruth Houbertz for discussing the unique specifics of the BFO-based, non-linear memristor compared to linear memristors, and Prof. Holger Boche, Prof. Gitta Kutyniok, and Adalbert Fono from TU München for pointing out the importance of developing analog hardware to realize computability of input-output maps of transcendental functions. I wish to thank the Federal Agency for Disruptive Innovation (SPRIN-D) for promoting this work as part of a validation order.

The data that supports the findings of this study are available within the article.